\documentclass[10pt,conference]{IEEEtran}
\IEEEoverridecommandlockouts

\usepackage[printonlyused]{acronym} 
\usepackage{float}
\usepackage[caption=false,font=normalsize,labelfont=sf,textfont=sf]{subfig}
\usepackage{graphicx}
\usepackage{xurl}
\usepackage{url}
\usepackage{amsmath,amssymb}
\usepackage{booktabs}
\usepackage{enumitem}
\usepackage{color}
\usepackage{xcolor}
\usepackage{cite}
\usepackage{threeparttable}

\definecolor{lightPurple2}{HTML}{9877BE}

\newcommand{\cmmnt}[1]{\ignorespaces}

\newif\ifcomment
\commenttrue

\ifcomment
    \newcounter{JOINumberOfComments}
    \stepcounter{JOINumberOfComments}
    
    \newcounter{AELNumberOfComments}
    \stepcounter{AELNumberOfComments}
    
    \newcounter{OzguNumberOfComments}
    \stepcounter{OzguNumberOfComments}
    
    \newcounter{DPNOTENumberOfComments}
    \stepcounter{DPNOTENumberOfComments}

      \newcounter{ARNOTENumberOfComments}
    \stepcounter{ARNOTENumberOfComments}

        \newcounter{PANumberOfComments}
    \stepcounter{PANumberOfComments}

\acrodef{3GPP}{3rd Generation Partnership Project}
\acrodef{2G}{Second Generation}
\acrodef{3G}{Third Generation}
\acrodef{4G}{Fourth Generation}
\acrodef{5G}{Fifth Generation}
\acrodef{APN}{Access Point Name}
\acrodef{AR}{Augmented reality}
\acrodef{BBU}{Baseband Unit}
\acrodef{BH}{backhaul}
\acrodef{BS}{Base Station}
\acrodef{BW}{bandwidth}
\acrodef{BLER}{block error rate}
\acrodef{CA}{Carrier aggregation}
\acrodef{CN}{Core Network}
\acrodef{CID}{Cell ID}
\acrodef{C-RAN}{Cloud-RAN}
\acrodef{CV}{Coefficient of Variation}
\acrodef{C-V2X}{Cellular \ac{V2X}}
\acrodef{CU}{Central Unit}
\acrodef{DD}{Decimal Degrees}
\acrodef{DL}{Downlink}
\acrodef{DM}{Demodulation}
\acrodef{DRL}{Deep \ac{RL}}
\acrodef{DU}{Distributed Unit}
\acrodef{EARFCN}{E-UTRA Absolute Radio Frequency Channel Number}
\acrodef{eMBB}{enhanced Mobile Broadband}
\acrodef{eNB}{E-UTRAN Node B}
\acrodef{ETSI}{European Telecommunications Standards Institute}
\acrodef{FH}{fronthaul}
\acrodef{Gb}{Gigabit}
\acrodef{GDPR}{General Data Protection Regulation}
\acrodef{GSM}{Global System for Mobile communication}
\acrodef{GSMA}{Global System for Mobile communications Association}
\acrodef{GPS}{Global Positioning System}
\acrodef{HO}{Handover}
\acrodef{HARQ}{hybrid automatic repeat request}
\acrodef{IMEI}{International Mobile Equipment Identity}
\acrodef{IMSI}{International Mobile Subscriber Identity}
\acrodef{IoT}{Internet of Things}
\acrodef{IS}{Indoor Static}
\acrodef{KDE}{Kernel Density Estimate}
\acrodef{KPI}{Key Performance Indicator}
\acrodef{LPWA}{Low-Power Wide-Area}
\acrodef{M2M}{Machine-to-Machine}
\acrodef{MCC}{Mobile Country Code}
\acrodef{MCS}{modulation and coding scheme}
\acrodef{MDP}{Markov Decision Process}
\acrodef{MEC}{Multi-Access Edge Computing}
\acrodef{M-IoT}{Multimedia IoT}
\acrodef{MH}{midhaul}
\acrodef{ML}{Machine Learning}
\acrodef{mmWave}{millimeter wave}
\acrodef{MME}{Mobility Management Entity}
\acrodef{mMTC}{massive Machine Type Communication}
\acrodef{MNC}{Mobile Network Code}
\acrodef{MNO}{Mobile Network Operator}
\acrodef{MVNO}{Mobile Virtual Network Operator}
\acrodef{MIMO}{Multiple-Input Multiple-Output}
\acrodef{MSC}{Message Sequence Chart}
\acrodef{MSE}{mean-squared-error}
\acrodef{NB-IoT}{Narrowband Internet of Things}
\acrodef{NDA}{Non-Disclosure Agreement}
\acrodef{NG-RAN}{Next Generation-Radio Access Network}
\acrodef{NR}{New Radio}
\acrodef{NR-ARFCN}{New Radio-Absolute Radio Frequency Channel Number}
\acrodef{NSA}{Non-Standalone}
\acrodef{NSPL}{National Statistics Postcode Lookup}
\acrodef{NSA}{Non-Standalone}
\acrodef{LoS}{Line of Sight}
\acrodef{LTE}{Long-Term Evolution}
\acrodef{LTE-M}{LTE for Machine-Type Communications}
\acrodef{OA}{Output Areas}
\acrodef{OD}{Outdoor Driving}
\acrodef{OFDM}{Orthogonal Frequency Division Multiplexing}
\acrodef{ONS}{Office for National Statistics}
\acrodef{OS}{Operating System}
\acrodef{OSD}{Outdoor Static Direct}
\acrodef{OSI}{Outdoor Static Inverse}
\acrodef{OT}{Outdoor Train}
\acrodef{OW}{Outdoor Walking}
\acrodef{PBCH}{Physical Broadcast Channel}
\acrodef{PCI}{Physical Cell ID}
\acrodef{RF}{Radio Frequency}
\acrodef{PSS}{Primary Synchronization Signal}
\acrodef{QCI}{QoS class Identifiers}
\acrodef{RAT}{Radio Access Technology}
\acrodef{RAN}{Radio Access Network}
\acrodef{RL}{Reinforcement Learning}
\acrodef{ROI}{Return on Investment}
\acrodef{RRC}{Radio Resource Control}
\acrodef{RRM}{radio resource management}
\acrodef{RRU}{Remote Radio Unit}
\acrodef{PDCP}{Packet Data Convergence Protocol}
\acrodef{RSRP}{Reference Signal Received Power}
\acrodef{RSRQ}{Reference Signal Received Quality}
\acrodef{RSU}{Roadside Unit}
\acrodef{RTT}{Round Trip Time}
\acrodef{RU}{Radio Unit}
\acrodef{QoS}{Quality of Service}
\acrodef{QCI}{QoS Class Identifier}
\acrodef{SA}{Standalone}
\acrodef{SIM}{subscriber identification module }
\acrodef{SCS}{Subcarrier Spacing}
\acrodef{SINR}{Signal to Interference and Noise Ratio}
\acrodef{SSB}{Synchronization Signal Block}
\acrodef{SS-RSRP}{Secondary Synchronization Reference Signal Received Power}
\acrodef{SSS}{Secondary Synchronization Signal}
\acrodef{SA}{Standalone}
\acrodef{SGSN}{Serving GPRS Support Node}
\acrodef{SGW}{Serving Gateway}
\acrodef{SDU}{Service Data Unit}
\acrodef{std}{Standard deviation}
\acrodef{PC}{Primary Component}
\acrodef{PCA}{Primary Component Analysis}
\acrodef{PRB}{physical resource block}
\acrodef{TAC}{Type Allocation Code}
\acrodef{TB}{transport block}
\acrodef{TDP}{thermal design power}
\acrodef{TTI}{Transmission Time Interval}
\acrodef{UE}{User Equipment}
\acrodef{UL}{Uplink}
\acrodef{VR}{Virtual Reality}
\acrodef{V2X}{Vehicle-to-Everything}
\acrodef{UTC}{Universal Time Coordinated}
\acrodef{URLLC}{ultra-reliable low-latency communication}
\acrodef{xDR}{eXtended Detail Record}

\begin{document}

\title{Energy-Efficient Task Computation at the Edge\\ for Vehicular Services}

\thispagestyle{plain}
\pagestyle{plain}

\author{\IEEEauthorblockN{Paniz Parastar\IEEEauthorrefmark{1}, Giuseppe Caso\IEEEauthorrefmark{2}, Jesus Alberto Omaña Iglesias\IEEEauthorrefmark{3}, Andra Lutu\IEEEauthorrefmark{3}, Özgü Alay\IEEEauthorrefmark{1}\IEEEauthorrefmark{2}
}
\IEEEauthorblockA{
\IEEEauthorrefmark{1}\textit{University of Oslo},
\IEEEauthorrefmark{2}\textit{Karlstad University},
\IEEEauthorrefmark{3}\textit{Telefonica Research}}

}

\maketitle

\begin{abstract}
Multi-Access Edge Computing (MEC) is a promising solution for providing the computational resources and low latency required by vehicular services, such as autonomous driving. It enables cars to offload computationally intensive tasks to nearby servers. Effective offloading involves determining when to offload tasks, selecting the appropriate MEC site, and efficiently allocating resources to ensure optimal performance. While car mobility poses significant challenges to guaranteeing reliable task completion, today we lack energy-efficient solutions to solve this problem, especially when considering 
real-world car mobility traces. In this paper, we begin by examining the mobility patterns of cars using data obtained from a leading mobile network operator in Europe. Based on the insights from this analysis, we design an optimization problem for task computation/offloading, considering both static and mobility scenarios. Our objective is to minimize the total energy consumption—both at the cars and the MEC nodes—while satisfying the latency requirements of various tasks. We evaluate our solution, based on multi-agent reinforcement learning, both in simulations as well as in a realistic setup that relies on datasets from the operator. Our solution shows a significant reduction of user dissatisfaction and task interruptions in both static and mobile scenarios, while achieving energy savings of 47\% (static) and 14\% (mobile) compared to state-of-the-art schemes. 

\end{abstract}

\section{Introduction}

\ac{V2X} services, such as car safety applications and assisted/autonomous driving, require the computation of tasks with strict latency and/or computational requirements \cite{5gpp2015, liu2019vec}. 
\ac{MEC} is a promising solution to meet these requirements, enabling \textit{task offloading}, i.e., tasks generated by vehicles may be transferred to nearby \ac{MEC} nodes for their computation \cite{hu2015etsi-mec, 5gaa2017}.  
This is essential to handle complex and resource-intensive tasks (e.g, object detection and recognition, data processing for anomaly detection) that vehicles themselves are unable to compute efficiently and/or quickly. 
In this context, effective task computation strategies encompass three decisions, i.e., (i) to offload a task to a \ac{MEC} node or compute it locally in the vehicle, (ii) if decide to offload, which \ac{MEC} node among the available ones to select, and (iii) the amount of computational resources to allocate on the selected MEC node, in order to satisfy task requirements (e.g., in terms of latency). 

\begin{figure}[t!]
    \centering
    \includegraphics[width=\columnwidth]{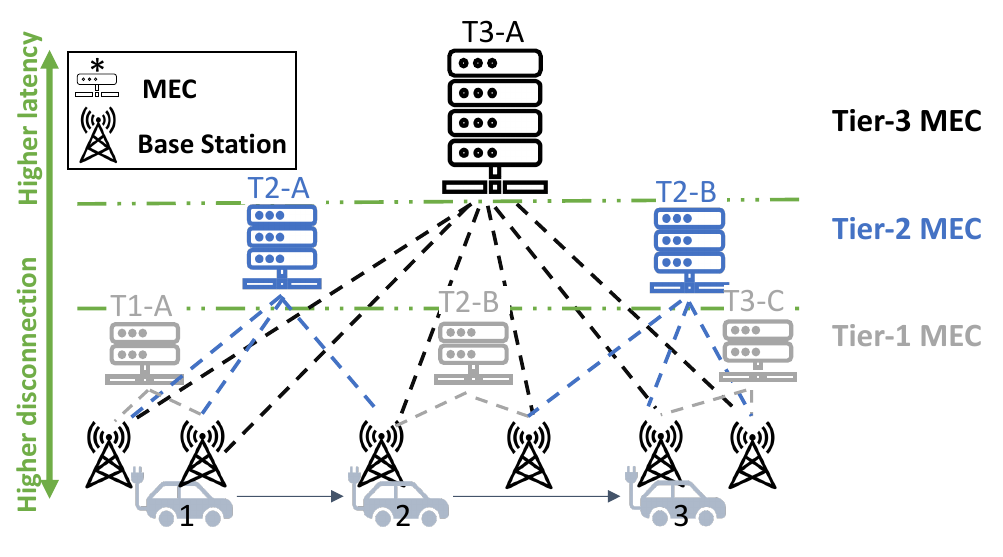}
    \caption{Multi-tier \ac{MEC} architecture for V2X services.}
    \label{fig:problem_statement}
    \vspace{-7mm}
\end{figure}

In parallel, the efficiency of a \textit{multi-tier} \ac{MEC} architecture to address service heterogeneity has been demonstrated~\cite{etsiV2XMEC2018, 5gaa2020c, Parastar2023}. 
As depicted on a high level in Fig.~\ref{fig:problem_statement}, higher tier \ac{MEC} nodes offer more computational resources with a larger coverage area, at the cost of higher latency. 
Prior studies \cite{Tong2016, Wang2019HetMec, Cartas2019}, including our previous work  \cite{Parastar2024}, have shown that a \textit{3-tier} MEC architecture provides a good tradeoff between latency and cost efficiency. 
This architecture aligns seamlessly with the disaggregated \ac{5G} and Beyond-\ac{5G} Radio Access Network (RAN) architecture, by strategically placing \ac{MEC} nodes at the physical locations of Distributed Units (DUs), Central Units (CUs), and Core instance sites. Motivated by these findings, we adopt a \textit{3-tier} MEC deployment in this work to cater  to the diverse demands of V2X services.

Vehicle/car mobility introduces significant challenges to task offloading as cars may lose access to their assigned \ac{MEC} nodes, requiring task reallocation. For instance, if a task of the car depicted in Fig. \ref{fig:problem_statement} is assigned to \textit{MEC T1-A} in Location 1, this \ac{MEC} node becomes inaccessible as the car moves to Location 2, requiring the task to be reassigned (if its computation is not finalized yet). On the other hand, assigning the task to \textit{MEC T2-A} allows to maintain connectivity in both locations. Ignoring car mobility can thus lead to sub-optimal task computation solutions. 

Studies that have considered user mobility in task computation strategies often neglect energy consumption and only focus on either optimizing latency or cost \cite{Saleem2020, Li2021, Dai2022, Zhang2024}, or only consider local (at the user end) energy without accounting for \ac{MEC} energy consumption \cite{Chen2022, Liu2022, Lai2024, Hua2024}. Additionally, most of these studies are based on simulations and either assume predefined mobility models, such as random walk \cite{Hu2021, Chen2022}, SMOOTH random \cite{Guan2022, Raza2022}, and Boundless Simulation Area (BSA) \cite{Hoang2020, Liu2022}, or require precise information for trajectory prediction, such as speed or movement direction at each decision time \cite{Lai2024, Cao2024}. In contrast, as detailed in the next sections, we investigate real-world car mobility patterns by analyzing the cars connected to the network of a major European \ac{MNO}. Based on this empirical investigation, we propose a task computation strategy 
relying on approximated cars' location information, i.e., the location of the \acp{BS} at which cars are connected to, which ultimately allows to avoid using personal and privacy-sensitive user location information.

In this paper, we first analyze cars' mobility patterns based on data captured from the \ac{MNO} under study. Then, exploiting the results of such analysis, we formulate an optimization problem for task computation, aimed at minimizing the total energy consumption (i.e., both at the cars and at the \ac{MEC} nodes) while meeting the latency demands of different tasks. We approach the solution to this problem by modeling it as a \ac{MDP}, which enables to handle the stochastic nature of car mobility as well as network dynamics. We propose a solution based on the multi-agent Proximal Policy Optimization (PPO) algorithm \cite{NEURIPS2022}, for both static and mobile scenarios. 
This approach enhances task computation efficiency and contributes to \ac{MEC} sustainable operations by reducing its energy consumption.

We summarize the main contributions as follows: 
\begin{itemize}[leftmargin=*, topsep=3pt] 
    \item We analyze cars' mobility patterns based on data captured in the network of a European \ac{MNO}, creating an hourly dataset that contains the \texttt{top20} \acp{BS} for each car, based on the connection duration. Our findings reveal that analyzing the \texttt{top20} \acp{BS} is sufficient to capture cars' mobility behaviors. In addition, we observe rather stable connectivity behaviors for cars over consecutive days (Sec. \ref{sec:empriical_mob}).
    \item To fully exploit heterogeneous computational resources, we consider a 3-tier \ac{MEC} platform, catering to various task requirements. Building on this platform, we formulate an optimization problem for solving task computation, aimed at minimizing energy consumption while meeting task latency requirements (Sec. \ref{sec:system_model}).
    \item Based on our car mobility analysis and on our system model, 
    we propose a task computation solution that adapts to both static and mobile scenarios, referred to as Latency-Aware PPO (LAPPO) in the static case and Mobility-Aware LAPPO (MALAPPO) in the mobile case. We evaluate LAPPO in a simulated setup and MALAPPO in a real-world setup that leverages the \ac{MNO}'s network topology. 
    Across scenarios, our approach decreases the number of cars for which task latency requirements are not met and task interruptions, over several baselines.   
    Moreover, it decreases the energy consumption by 47\% (static) and 14\% (mobile). These results show the efficiency of our approach across different network topologies and mobility scenarios (Sections \ref{sec:methodology} and \ref{sec:eval_main}).
\end{itemize}

%


\section{Related Work}

Efficient task computation/offloading, especially in the presence of high dynamics due to car mobility, pose a significant challenge in \ac{MEC} systems.
Several works (e.g., \cite{Hoang2020, Dai2022, Guan2022, Hua2024}) have explored user mobility in the task computation decisions, mainly choosing between local computation vs. offloading to a nearby \ac{MEC} node (co-located with \ac{BS} or \ac{RSU}) vs. offloading to the cloud. However, these works focus on deciding between local devices, edge, or cloud nodes, overlooking the resource allocation problem, i.e., how to distribute computational resources among tasks in systems with multiple mobile users and edge/cloud nodes. 
Other works filled the above gap by determining the amount of resources to be allocated to various tasks (e.g., \cite{Din2019, Yang2019, Zhang2020, Zhan2020, Li2021, Cao2024}). Nevertheless, the heterogeneous computational capabilities of edge nodes, along with the need for selecting the best suited node among the available ones 
add to the complexity of the problem. The work in \cite{Peng2019, Saleem2020, Hu2021, Chen2022, Lai2024} moves a step beyond by simultaneously optimizing edge node selection and resource allocation decisions assuming to know users' location, but the inequality of the computational resources needed for the tasks was not considered, thus leading to inefficient resource allocation and high energy consumption \cite{Cao2024}. A subset of works addressed this gap and studied unequal edge resource allocation to maximize the total system revenue \cite{Zhang2024}, to minimize the latency experienced by users \cite{Wang2023}, or to maximize computation efficiency \cite{Raza2022}. 


Unlike several works that consider homogeneous \ac{MEC} nodes co-located with every \ac{BS}/\ac{RSU} (e.g., \cite{Tan2018, Wang2019, Hoang2020, Hu2021, Raza2022, Dai2022, Lai2024, Hua2024, Zhang2024, Cao2024}), we build our task computation scheme on a multi-tier \ac{MEC} architecture, where higher tiers have larger computational capacity while lower tiers have lower capacity but also lower latency towards the users.
Additionally, considering the importance of carbon footprint and energy consumption, we incorporate energy consumption as our objective, standing apart from works ignoring energy optimization (e.g., \cite{Peng2019, Wang2019, Li2021, Wang2023, Zhang2024}) or focusing on energy savings on local devices (e.g., \cite{Din2019, Hoang2020, Saleem2020, Zhan2020, Hu2021, Dai2022, Raza2022, Guan2022, Chen2022, Liu2022, Lai2024, Hua2024}).
Finally, we observe that many works either use statistical mobility models in their evaluations (e.g., \cite{Zhan2020, Tan2018, Peng2019, Hu2021, Chen2022, Enayet2018, Guan2022, Raza2022, Hoang2020, Liu2022}) or assume to have access to precise information on user location, moving speed, or movement direction (e.g., \cite{Peng2019, Lai2024, Cao2024}), which may not be available due to privacy concerns. We instead use a real-world dataset to capture high-level mobility patterns, accessing only to the location of the \acp{BS} to which cars connect over time, and  using it as a proxy for their actual location.
\section{Empirical study on car mobility}\label{sec:empriical_mob}
\begin{figure*}[hbt!]
\subfloat[]{
\includegraphics[width=0.32\textwidth]{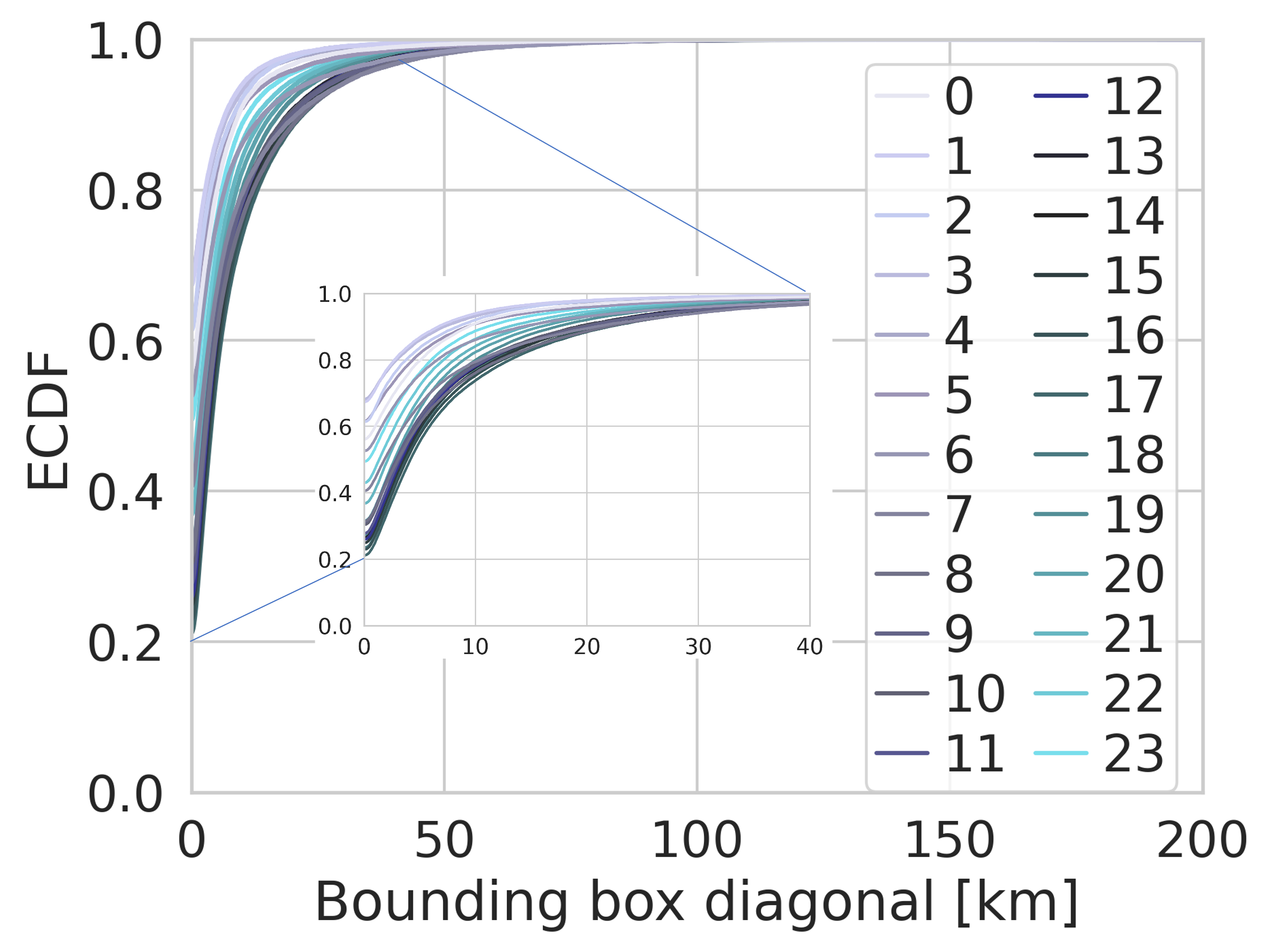}
    \label{fig:mob_bbdiagonal}
}
 \subfloat[]{
\includegraphics[width=0.32\textwidth]{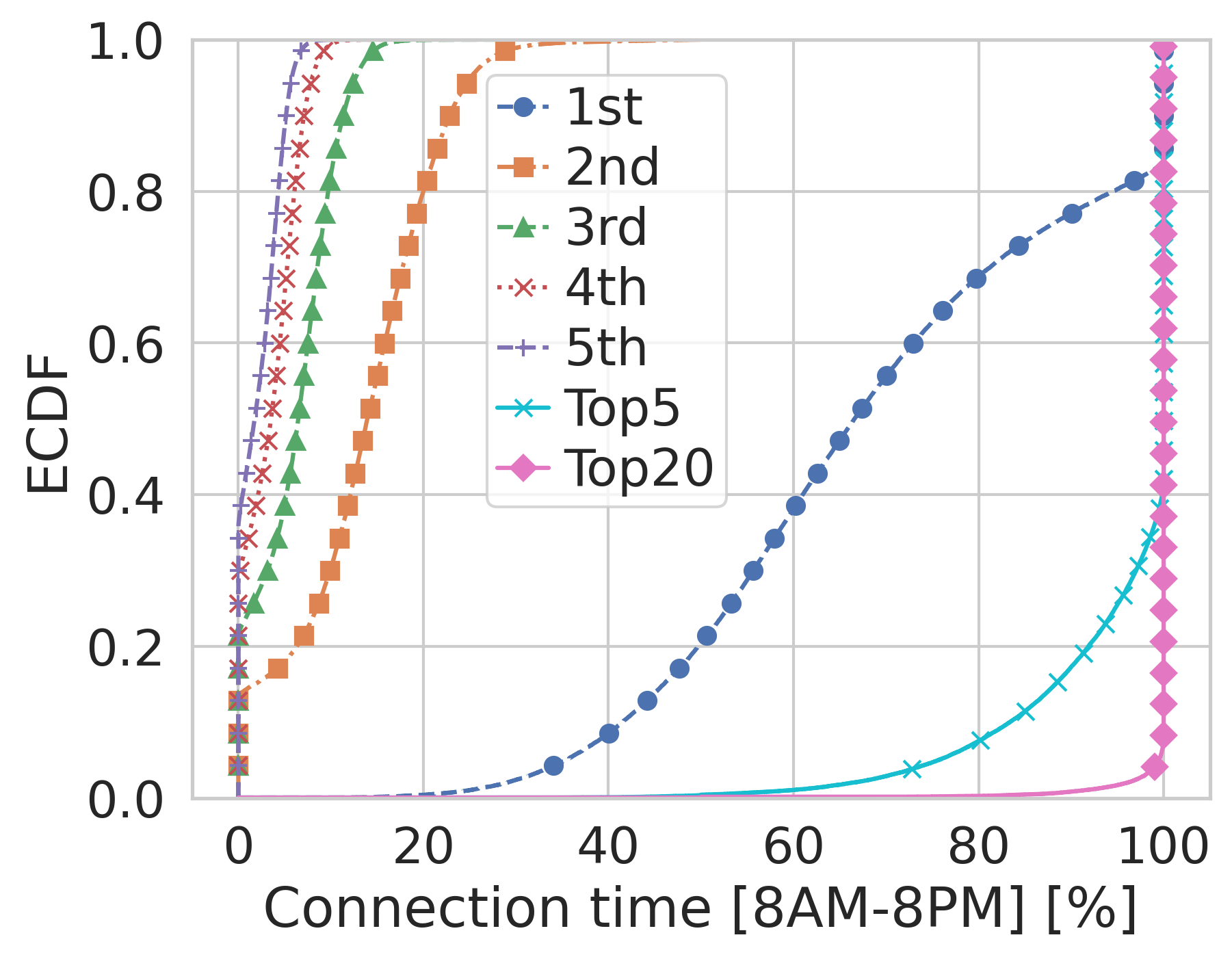}
    \label{fig:mob_connectiontime}
}
 \subfloat[]{
\includegraphics[width=0.32\textwidth]{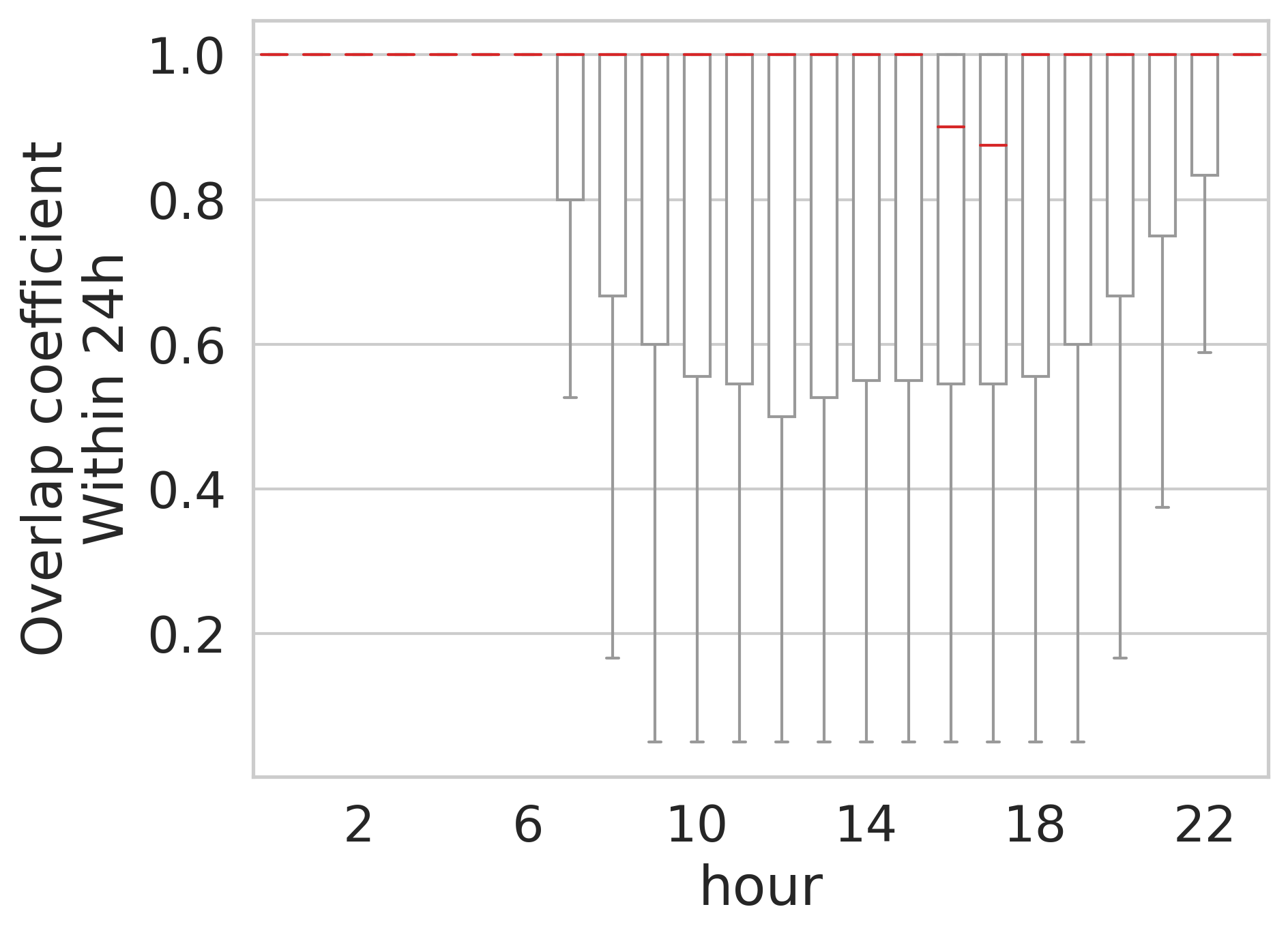}
    \label{fig:mob_overlap}
}
\caption{Mobility characteristics of cars in our dataset. (a) Per-hour ECDF of the bounding box diagonal traveled by cars. (b) Percentage of cars' connection time to the \texttt{top20} and \texttt{top5} (total and per BS) BSs, between 8 AM and 8 PM. (c) Per-hour overlap coefficient (statistics over cars).}        
\vspace{-7mm}
\label{fig:mob_exploration}
\end{figure*}

In this section, we explore real-world car mobility patterns by analyzing the data we captured on the network of a major \ac{MNO} in Europe. We build a car-level hourly dataset where we store the \texttt{top20} \acp{BS}. These are the 20 \acp{BS} where each car spends the longest amount of time connected, within an hour. 
We process three weekdays of data in May 2024, which includes about 60,000 cars that maintain stable network connections (i.e., cars connected to the network every day from 8 AM to 8 PM) during the analysis period, ensuring a robust and reliable sample for our study.

We first explore the space traveled by cars over one hour. To do so, we compute the bounding box diagonal (i.e., the Euclidean distance between minimum and maximum coordinates) \cite{Becker2013} per car per hour. Then, we calculate the median over days and plot the empirical cumulative distribution function (ECDF) for different hours in Fig.  \ref{fig:mob_bbdiagonal}. Our results show that car movement is lower during the late night and early morning hours (as lines with light colors indicate in the figure), with 95\% of cars traveling less than 15 km between 12 AM and 5 AM. We thus limit our next analyses between 8 AM and 8 PM to reduce the nighttime effect. In Fig. \ref{fig:mob_bbdiagonal}, we also observe that, during a day, 95\% of cars drive less than 30 km (0.95-quantile is 29.9 km from 6 PM to 7 PM). 

Fig. \ref{fig:mob_connectiontime} shows the ECDF of the percentage of connection time to the \texttt{top20} \acp{BS} and \texttt{top5} \acp{BS} (\texttt{top5} \acp{BS} also separately). We find that 92\% of cars only connect to 20 \acp{BS}, and the median connection time to the \texttt{top20} \acp{BS} is ${\sim}99\%$ for the remaining cars. By analyzing the \texttt{top20} \acp{BS}, we can thus capture most of the mobility patterns, even for those cars connecting to more than 20 \acp{BS}. By exploring the connection time to the \texttt{top5} \acp{BS}, we notice that 60\% of cars only connect to their \texttt{top5} \acp{BS}, with the majority of their time on the \texttt{top1} BS (the median of connection time on this BS is 66\%). We also find that 16\% of cars connect to only one \ac{BS}.

Next, to understand the daily connection stability of cars, we analyze the similarity between the set of \acp{BS} cars connect to over time. In particular, we compute the overlap coefficient ($\mathrm{oc}$) between the set of \acp{BS} used during a particular hour in a specific day and the set of \acp{BS} used during the previous 24 hours. Hence, for each car $c$ in our dataset, we define $\mathcal{A}_{c}^{d,h}$ 
as the set of \acp{BS} that car $c$ connects to during hour $h$ on day $d$, while $\mathcal{A}_{c}^{d-1:d,h}$ 
is the set of \acp{BS} that the same car $c$ connects to between hour $h$ on day $d-1$ and hour $h$ on day $d$, i.e.,
\begin{equation}
\vspace{-2mm}
\mathcal{A}_{c}^{d-1:d,h} = \bigcup\limits_{h^{\prime}=h}^{23} \mathcal{A}_{c}^{d-1, h^{\prime}} \cup \bigcup\limits_{h^{\prime}=0}^{h-1} \mathcal{A}_{c}^{d, h^{\prime}}. 
\end{equation}

The overlap coefficient is then computed as \eqref{eq:overlap}, where we substitute $\mathcal{A}_{c}^{d,h}$ with $\mathcal{Q}$ and $\mathcal{A}_{c}^{d-1:d,h}$ with $\mathcal{V}$ for simplicity:

\begin{equation}
    \mathrm{oc}(\mathcal{Q}, \mathcal{V}) = \frac{|\mathcal{Q} \cap  \mathcal{V}|}{\text{min}(|\mathcal{Q}|, |\mathcal{V}|)}.
    \label{eq:overlap}
    \vspace{-1mm}
\end{equation}

Fig. \ref{fig:mob_overlap} shows the boxplots of hourly overlap coefficients (i.e., we compute the median per car over the analysis period). We observe that the median of overlap coefficients is 1 except for hours 16 and 17 (4 PM and 5 PM). This indicates that, even though cars move, the \ac{BS} set cars connect to within an hour is typically a subset of their daily associated \ac{BS} set, except for a slight variation during evening peak hours. Therefore, cars exhibit stable connectivity behaviors over consecutive days and frequently reconnect to the same set of \acp{BS}.

\begin{table}[h!]
\vspace{-4mm}
\begin{threeparttable}
\setlength{\tabcolsep}{2.7pt}
\renewcommand{\arraystretch}{1}
\caption{\small Daily Mobility Clustering Results.\tnote{*}}
\small
\label{tab:ccar_clustering}
\begin{tabular}{@{}|l|l|l|l|l|l|l|l|l|l|l|l|l|l|@{}}
\hline
\textbf{\rotatebox[origin=c]{90}{\begin{tabular}[c]{@{}l@{}}Cluster\\(\% of cars\\in the cluster)\end{tabular}}} & \textbf{\rotatebox[origin=c]{90}{\begin{tabular}[c]{@{}l@{}}Gyration \\{[}meter{]}\end{tabular}}} & 
\textbf{\rotatebox[origin=c]{90}{\begin{tabular}[c]{@{}l@{}}Total connection\\ time {[}hour{]}\end{tabular}}} & \textbf{\rotatebox[origin=c]{90}{\texttt{Top20}\tnote{$\dagger$}}} & \textbf{\rotatebox[origin=c]{90}{\texttt{1st}}} & \textbf{\rotatebox[origin=c]{90}{\texttt{2nd}}} & \textbf{\rotatebox[origin=c]{90}{\texttt{3rd}}} & \textbf{\rotatebox[origin=c]{90}{\texttt{4th}}} & \textbf{\rotatebox[origin=c]{90}{\texttt{5th}}} & \textbf{\rotatebox[origin=c]{90}{\texttt{6th}}} & \textbf{\rotatebox[origin=c]{90}{\texttt{7th}}} & \textbf{\rotatebox[origin=c]{90}{\texttt{8th}}} & \textbf{\rotatebox[origin=c]{90}{\texttt{9th}}} & \textbf{\rotatebox[origin=c]{90}{\texttt{10th}}} \\ \cline{1-14}
\begin{tabular}[c]{@{}l@{}}high mobile\\ (40\%)\end{tabular} & 4538              & 21.3                                                                                 & 99             & 48         & 20         & 9          & 4          & 2          & 1          & 1          & 1          & 0          & 0           \\  \cline{1-14}
\begin{tabular}[c]{@{}l@{}}low mobile\\ (60\%)\end{tabular}  & 461               & 21                                                                                   & 100            & 90         & 6          & 0          & 0          & 0          & 0          & 0          & 0          & 0          & 0           \\ \cline{1-14}
\end{tabular}
\begin{tablenotes}\footnotesize
\item[*] The values are the median of the features within each cluster.
\item[$\dagger$] {The last 11 columns denote the percentage [\%] of the total connection time to the \texttt{top20} \acp{BS} and for each \texttt{top10} \ac{BS} separately. We omit \acp{BS} from \texttt{11} to \texttt{20} as their values are 0.}
\end{tablenotes}
\end{threeparttable}
\vspace{-3mm}
\end{table}

We conclude this analysis by clustering cars based on their daily mobility pattern. To do so, we consider mobility features as gyration, total connection time, and the percentage of connection time to \texttt{top20} \acp{BS} in total and separately (i.e., 23 features altogether). Gyration is a mobility metric that models the distance that cars travel during the day in terms of how far they are located from the centroid \cite{Gonzalez2008}. 
We compute it as $\sqrt{\frac{1}{N} \sum_{i=1}^N{(t_i I_i - I_{cm})}^2}$, 
where $N$ is the total number of \acp{BS} visited during a day, and $t_i$ denotes the amount of time that the car spent on \ac{BS} $i$, this latter being deployed at location $I_i$. $I_{cm}$ represents the location of the centroid, which is computed as $\frac{1}{N} \sum_{i=1}^N{t_i I_i}$. We adopt the density-based spatial clustering of applications with noise (DBSCAN) algorithm \cite{dbscan} to cluster cars based on their mobility behaviors. We discover two mobility clusters, which we identify as \textit{high mobile} and \textit{low mobile}. The \textit{high mobile} cluster includes 40\% and the \textit{low mobile} includes 60\% of the cars. We report the median of features per cluster in Table 
\ref{tab:ccar_clustering}. 

Our analysis in this section highlights distinct patterns in car mobility behaviors, which necessitate tailored approaches for task computation, aiming to adapt to varying mobility conditions, from static to high mobility. 
As further detailed in the next sections, we first propose the 
LAPPO scheme for the static scenario, i.e., for handling the majority of cars during late night and early morning hours that, as reported above, remain connected to the same \ac{BS} for most of the time. 
We then address the mobile scenario by extending LAPPO to MALAPPO, while also leveraging the clustering results obtained in this section.  
We show that this extension is key for managing the dynamic car connectivity during peak hours, ensuring a low number of transitions between \ac{MEC} nodes and low energy consumption.


\section{System model}\label{sec:system_model}

\sloppy Our system model includes a set of car
$\mathcal{C} = \{c\text{ }|\text{ }1 \leq c \leq C\}$, a set of \ac{MEC} nodes $\mathcal{M} =  \{m\text{ }|\text{ }1 \leq m \leq M\}$, and a set of \acp{BS} $\mathcal{B} = \{b\text{ }|\text{ }1 \leq b \leq B\}$. 

We assume that each \ac{BS} is connected to a limited set of \ac{MEC} nodes, depending on the underlying network infrastructure topology. Therefore, when a car connects to a \ac{BS}, it can only access to the \ac{MEC} nodes associated with that particular \ac{BS}. Due to mobility, cars connect to different BSs over time, and thus access to different sets of MEC nodes. 

Given a time instant $t$, each \ac{MEC} node is defined by the tuple $\{lat_m, lon_m, cpu_m(t), trans_m(t)\}$, where 
$lat_m$ and $lon_m$ identify the location of \ac{MEC} node $m$ (not depending on $t$ as we assume that \ac{MEC} nodes do not move), while $ cpu_m(t)$ and $trans_m (t)$ denote the available computational capacity and data transmission rate of \ac{MEC} node $m$ at time $t$, 
both further detailed later in the section. Similarly, each car $c$ at time $t$ is defined by the tuple $\{b(t), lat_b(t), lon_b(t), mob_c(t), cpu_c(t), E_c (t), \mathcal{M}_c(t), k_c(t)\}$, where $lat_b(t)$ and $lon_b(t)$ identify the location of the \ac{BS} at which the car is connected at time $t$, i.e., $b(t)$.\footnote{We assume that privacy-sensitive information about car location is not available, and thus we approximate the location of a car to the location of the BS at which the car is connected, at each time $t$.} Moreover, $mob_c (t)$ is the car mobility group indicating the daily mobility behavior of the car (as defined in Sec.  \ref{sec:empriical_mob}), $cpu_c (t)$ is the computational capacity available in the car, $E_c(t)$ is the car battery energy level, and $\mathcal{M}_c(t) \subset \mathcal{M}$ is the set of \ac{MEC} nodes accessible to the car, which depends on $b(t)$.\footnote{For notation simplicity, the reference to the specific BS enabling access to a specific set of \ac{MEC} nodes is omitted when defining $\mathcal{M}_c(t)$, due to the univocal mapping between a car and a BS at each time $t$.} Finally, $k_c(t)$ is the task generated by the car $c$ at time $t$, for which a strategy must be decided (e.g., local computation or offloading to a \ac{MEC} node). 

At time $t$, assuming that a car can generate up to one task, there will be a set of tasks $\mathcal{K}(t) = \{k \text{ }|\text{ }1 \leq k \leq K\}$ needed to be computed.\footnote{For notation simplicity, the reference to the car generating each task is omitted when defining the set $\mathcal{K}(t)$, due to the univocal mapping between a car and a task at each time $t$.} Each task $k \in \mathcal{K}(t)$ is defined by a tuple $\{data_k, cl_k, l^{max}_k\}$, where $data_k$ is the size of the task (e.g., in bytes), $cl_k$ is the number of CPU cycles needed to process one bit, and $l^{max}_k$ is the maximum latency tolerable by the task. Following previous work (e.g., \cite{Hu2021, Li2021Dynamic, Gan2022}), we consider the required computation intensity for a task as $ci_k = data_k \cdot cl_k$. 
Moreover, as typical in the literature (e.g, \cite{Dai2022, Zhang2024, Hua2024, Zhan2020}), we assume that tasks are atomic, i.e., they are executed in one place and cannot be split.



Assuming that a car $c \in \mathcal{C}$ has a task $k \in \mathcal{K}(t) $ to be computed, let $x_k^m (t) \in \{0, 1\}$ denote whether task $k$ is offloaded ($x_k^m (t) = 1$) or not ($x_k^m (t) = 0$) on \ac{MEC} node $m \in \mathcal{M}_c(t)$. 
Additionally, we define $y_k(t) = 1 - \sum_{m \in \mathcal{M}_c(t)} x_k^{m}(t)$, 
where $y_k(t) = 1$ represents local computation and thus, by definition, $y_k(t) = 0$ if the task is offloaded to a MEC node.  

Given the above notation, we denote the latency experienced by task $k$ as $l_k(t)$ and the energy consumption for its computation as $e_k(t)$. 
$l_k(t)$ can be expressed as follows:
\begin{equation}
    \label{eq:l_k}
    l_k(t) = y_k(t) \cdot l_k^\mathrm{loc}(t) + \sum_m x_k^m(t) \cdot l_k^m(t)
\vspace{-2mm}
\end{equation}
where $l_k^{\mathrm{loc}}(t)$ is the local (car) latency and $l_k^m(t)$ is the latency due to the offloading of the task to \ac{MEC} node $m$. 
$e_k(t)$ is a weighted sum of the energy consumed by the car, $e^c_k(t)$, and the energy consumed by the MEC node if the car has decided to offload its task, $e^{\mathrm{off}}_k(t)$, as follows: 
\begin{equation}
\label{eq:e_k}
    e_k(t) = e_k^{c}(t) + \alpha \cdot e_k^{\mathrm{off}}(t), 
\vspace{-2mm}
\end{equation}
where $\alpha$ 
balances the importance of local energy relative to MEC energy consumption, and can be adjusted based on how crucial the local energy consumption is. 
If $\alpha \simeq 0$, the goal reduces to optimizing the local energy consumption, also reflecting possible additional costs in charging car batteries. When the car's battery life is less of a concern, we can use local computational resources more often and decrease the overall energy consumption, also considering the \ac{MEC} system side. Moreover, $e^c_k(t)$ and $e^{\mathrm{off}}_k(t)$ are evaluated as follows:
\begin{equation}
\label{eq:e_loc_off}
    \begin{cases}
e^{c}_k(t) &= y_k(t) \cdot e_k^{c,\mathrm{loc}}(t) + (1 - y_k(t)) \cdot e_k^{c, \mathrm{off}}(t)\\
e^{\mathrm{off}}_k(t) &= \sum_m x_k^m(t) \cdot e_k^{m}(t)
\end{cases}
\vspace{-2mm}
\end{equation}

On the one hand, $e^c_k(t)$ includes the energy consumed by the car for either local computation, $e_k^{c,\mathrm{loc}}(t)$, or for task offloading, $e_k^{c, \mathrm{off}}(t)$. On the other hand, $e_k^{\mathrm{off}}(t)$ is the energy consumed in the MEC system if the car has decided to offload its task to a MEC node $m$. Moreover, $E_c(t+1) = E_c(t) - e_k^{c}(t)$, so to reflect the energy consumed locally for computing or offloading the task.

In the next subsections, we detail latency and energy components involved in task computation locally vs. at \ac{MEC}.

\subsection{Latency and Energy with Local Task Computation}
Assume that car $c$ decides to locally compute its task, we evaluate $l_k^{\mathrm{loc}}(t)$ as follows:
\begin{equation}
\label{eq:l_loc}
l_k^\mathrm{loc}(t) = \frac{ci_{k}}{f_{k}(t) \cdot cpu_c (t)},
\vspace{-2mm}
\end{equation} 
where $f_{k}(t) \in (0,1]$ is the CPU frequency ratio allocated to the task.
As for the energy, we adopt a widely used model for energy consumption (used, for example, in \cite{Guo2016, Wang2020, Hu2021}), so that $e_k^{c,\mathrm{loc}}(t)$ is as follows: 
\begin{equation}
e_k^{c, \mathrm{loc}}(t) = ci_{k} \cdot \kappa \cdot [f_{k}(t) \cdot cpu_c(t)]^2,  
\vspace{-2mm}
\end{equation}
where, $\kappa = 10^{-11}$ is the effective switched capacitance depending on the chip architecture \cite{Guo2016}.

\subsection{Latency and Energy with Task 
Computation at \ac{MEC}}
Assume that car $c$ decides to offload its task to MEC node $m \in \mathcal{M}_c(t)$. We calculate $l^{m}_k(t)$ as the sum of communication ($l^{m, \mathrm{comm}}_{k}$) and computation ($l^{m, \mathrm{comp}}_{k}$) latencies, $e^{c, \mathrm{off}}_k(t)$, and $e^{m}_k(t)$ as follows, where  
we neglect the downlink communication latency to the car, assuming that the task execution result is minimal in size \cite{Zhan2020, Zhou2019}: 
\begin{equation}
\vspace{-1mm}
\begin{cases}
l^{m}_k(t) = l^{m, \mathrm{comm}}_{k} (t) + l^{m, \mathrm{comp}}_{k} (t)\\
     l^{m, \mathrm{comm}}_{k} (t)= l^{\mathrm{air}}_{k}(t) + l^{m, \mathrm{ts}}_k(t) + l^{m, \mathrm{tx}}_k(t)
    \end{cases}
\vspace{-2mm}
\label{eq:l_off}
\end{equation}

\begin{equation}
    \begin{cases}
        e^{c, \mathrm{off}}_k(t) = e^{\mathrm{air}}_k(t)\\
    e^{m}_k(t) = e^{m, \mathrm{tx}}_k(t) + e^{m, \mathrm{comp}}_{k} (t)
    \end{cases}
    \label{eq:e_off}
    \vspace{-1mm}
\end{equation}


\noindent We now detail latency and energy components in \eqref{eq:l_off} and \eqref{eq:e_off}:

\noindent \textbf{Over-the-air latency and energy} ($l^{\mathrm{air}}_{k}(t)$, $e^{\mathrm{air}}_{k}(t)$) -- $l^{\mathrm{air}}_{k}(t)$ is the time it takes to successfully deliver a task from the car to the \ac{BS} and ranges from 0.35 to 1.78 ms, which we derived following the specifications in \cite{ITU-M2499}. 
Then, $e^{\mathrm{air}}_k(t) = l^{\mathrm{air}}_{k}(t) \cdot p^{\mathrm{tx}}_c$, where $p^{\mathrm{tx}}_c$ is the car transmission power.

\noindent \textbf{Transport latency} ($l^{m, \mathrm{ts}}_k(t)$) -- This is the propagation time between the \ac{BS} to which the car is connected and the selected \ac{MEC} node. It depends on link length and propagation speed on the link. 
In order to model this component realistically, we use a liner regression model derived from real-world traces captured on the network of the \ac{MNO} under study, so that $l^{m, \mathrm{ts}}_k(t) = 0.014 \cdot \mathrm{dist}_{b, m}(t) + 1.225$, where $\mathrm{dist}_{b, m}(t)$ is the Euclidean distance between the \ac{BS}-\ac{MEC} pair used at time $t$, which is possible to calculate assuming to have information on the position of BSs and MEC nodes. 

\noindent \textbf{Transmission latency and energy} ($l^{m, \mathrm{tx}}_{k}(t)$, $e^{m, \mathrm{tx}}_{k}(t)$) -- $l^{m, \mathrm{tx}}_{k}(t)$ is the time it takes to transmit a task  onto the physical medium, i.e., $l^{m, \mathrm{tx}}_{k}(t)=\frac{data_{k}}{trans_{m} (t)}$. As mentioned above, $trans_{m}(t)$ is the transmission rate of the MEC node, which depends on how many cars are connected to it at time $t$. Assuming a maximum rate of $R_m$, $trans_{m}(t)$ is evaluated as the ratio between $R_m$ and the number of cars connected to MEC node $m$ at time $t$. 
The energy consumption is then $e^{m, \mathrm{tx}}_k (t) = l^{m, \mathrm{tx}}_{k} (t) \cdot p^{\mathrm{tx}}_b$, where $p^{\mathrm{tx}}_b$ is the transmission power of the \ac{BS} transmitting the task to the MEC node.

\noindent \textbf{\ac{MEC} computation latency and energy} ($l^{m, \mathrm{comp}}_{k} (t)$, $e^{m, \mathrm{comp}}_{k} (t)$) -- These components depend on the task computation intensity and the resources allocated in the selected \ac{MEC} node for its computation. 
We adopt a common formula for the latency, so that $l^{m, \mathrm{comp}}_{k} (t) = \frac{ci_{k}}{f_{k}(t) \cdot cpu_m (t)}$, where $f_{k} (t) \cdot cpu_m (t)$ is the amount of computational resources allocated to the task (e.g., see \cite{Li2021, Gan2022}). We assume that the \ac{MEC} node supports multi-processing, allowing it to execute multiple tasks concurrently. It is thus feasible to allocate a fraction of the resources to the task (i.e., $0 < f_{k} (t) \leq 1$). Assuming a maximum capacity of $\mathrm{CPU}_m$, $cpu_{m}(t) \leq \mathrm{CPU}_m$ is the computational capacity available at the MEC node $m$ at time $t$, which depends on the total requested computational resources of all users at that time. As for the energy, similar to the local computation, $e^{m, \mathrm{comp}}_{k} (t) = ci_{k} \cdot \kappa \cdot [f_{k} (t) \cdot cpu_m (t)]^2$.


\setlength{\belowdisplayskip}{2pt} \setlength{\belowdisplayshortskip}{2pt}
\setlength{\abovedisplayskip}{2pt} \setlength{\abovedisplayshortskip}{2pt}

\subsection{Problem Formulation} \label{sec:opt_model}
We aim at formulating an optimal task computation strategy, where cars with tasks to be computed apply a policy for deciding on computing the task locally or offloading its computation to one of the available MEC nodes. The goal is to minimize the overall energy consumption (at both cars and MEC nodes) over time, while satisfying the latency constraints of the tasks. 
Therefore, we formulate this problem as follows: 

\vspace{-4mm}
\begin{equation}\label{eq:obj}
\vspace{-1mm}
    Min \; \sum_t \sum_k e_k(t) 
\end{equation}

\begin{subequations}
\begin{align}
\label{eq:c_latency}
    &l_k(t) \leq l_k^{max}  \quad \forall k, t\\
\label{eq:c_car_capacity}
    &f_{k}(t) \cdot y_k(t) \leq 1  \quad \forall k, t\\
\label{eq:c_mec_capacity}
    &\sum_k f_{k}(t) \cdot x^m_k(t) \leq 1  \quad \forall m, t\\
        \label{eq:c_energy}
    &e^c_k(t) \leq E_c(t) \quad \forall c, t\\
\label{eq:c_car_1mec}
    &y_k(t) + \sum_m x_k^m(t) = 1  \quad \forall k, t \\
    \label{eq:x_binary}
    &x_k^m(t) \in \{0, 1\}  \quad \forall k, m, t
\end{align}
\end{subequations}

The objective function is in \eqref{eq:obj}, where we aim at minimizing the total energy consumption at cars and \ac{MEC} nodes over time. Constraint \eqref{eq:c_latency} takes into account the latency requirement of each task at each time, while \eqref{eq:c_car_capacity} and \eqref{eq:c_mec_capacity} ensure car and \ac{MEC} available computational capacities are not exceeded. Constraint \eqref{eq:c_energy} indicates that the energy consumed locally cannot exceed the battery energy level, while \eqref{eq:c_car_1mec} and \eqref{eq:x_binary} indicate that a car can offload its task to only one \ac{MEC} at time $t$ or compute it locally.

In the next section, we propose a solution based on Multi-Agent Reinforcement Learning (MARL) to solve this problem, assuming partial information available at each car.
\section{Proposed MARL-based Solution}\label{sec:methodology}

In stochastic environments, a decision-making problem as the one formulated in the previous section can be modeled as a MDP, since the next system state depends only on the current state and the actions performed by a decision-making agent. 
Typically, an MDP is defined by a tuple $\{\mathcal{S}, \mathcal{A} , p , r\}$ where $\mathcal{S}$ is a finite set of states, $\mathcal{A}$ is a finite set of actions, $p$ is a transition probability from a state $s \in \mathcal{S}$ to a state $s' \in \mathcal{S}$ after executing an action $a \in \mathcal{A}$, and $r$ is the immediate reward obtained when action $a$ is performed. The agent goal of is to find an optimal policy, i.e., a mapping from the state space to the action space, aiming to maximize the reward function over time. In a multi-agent, decentralized, and partially observable \ac{MDP}, there are $N$ agents and each agent $n$ only accesses to local observations $o_n$ that partially represent the state $s$. By considering their $o_n$, the goal of the agents is to jointly optimize a discounted accumulated reward $\mathbb{E}_{A^t,s^t} \sum_t \gamma^t r(s^t, A^t)$, where $A^t = (a^t_1,...,a^t_N)$ is the joint action of agents at time $t$, and $\gamma$ is a discount factor balancing the importance between immediate and future rewards.


RL algorithms are widely used to address \acp{MDP}, where an agent can learn their optimal policies by interacting with the environment. Among RL methods, we use the Proximal Policy Optimization (PPO) algorithm, due to its proven good performance in multi-agent scenarios \cite{NEURIPS2022}. 
PPO is 
based on the actor-critic architecture, where the critic network estimates the state-value function while the actor network optimizes the policy based on the estimated state-value function \cite{schulman2017ppo}.

\subsection{Latency-Aware PPO (LAPPO)}
We first design a solution, named LAPPO, where we adapt the multi-agent PPO (MAPPO) algorithm \cite{NEURIPS2022} to our scenario. MAPPO leverages centralized training with decentralized execution, enabling agents to operate efficiently in a mixed environment. During training, a centralized critic network takes the global state, i.e., a comprehensive representation of the entire environment, as the input for estimating the accumulated reward to optimize energy consumption. During execution, each agent independently derives its policy from its own actor network, based on local state observations, enabling competition for resource allocation. 
In our problem, cars are the agents, which observe their own state and take actions accordingly. State, action, and reward are defined as follows: 

\textbf{The state space.}
At any time $t$, the local state observation includes the status of the MEC nodes available to the car. As described in Sec. \ref{sec:system_model}, we assume a car can only connect to a limited set of \ac{MEC} nodes based on its \ac{BS}. Hence, we first add this information to the car's state observation through the binary vector $\{\mathrm{map}_1, \mathrm{map}_2, ..., \mathrm{map}_M\}$, where $\mathrm{map}_m = 1 \iff m \in \mathcal{M}_c(t)$. Then, we also include $l^m_k(t)$ and $cpu_m(t)$ for all \ac{MEC} nodes in $\mathcal{M}_c(t)$ (i.e., status of available MEC nodes), $cpu_c(t)$ and $E_c(t)$ (i.e., car status), and $ci_{k}$, $data_{k}$, and $l^{\mathrm{max}}_{k}$ (i.e., task information).

\textbf{The action space.}
The action space is represented by the task computation decision in the range of [0, $M$] and by the selection of a value for $f_k$ in the range of (0, 1]. At each $t$, each car has two possible choices: local execution or offloading its task to the one of \ac{MEC} nodes. For local execution, we set the value of task computation decision as 0 while the values in the range [1, $M$] denotes the selected \ac{MEC} node. As explained in Sec. \ref{sec:system_model}, $f_k$ is the CPU frequency ratio allocated to task $k$.   

\textbf{The reward function.}
The reward function is representative of the original objective function in \eqref{eq:obj}, as follows:
\begin{equation}
    r_c(t)=
    \begin{cases}
    -penalty, & m \notin \mathcal{M}_c \;\; \mathrm{or} \;\; l_k > l^{\mathrm{max}}_{k} \;\; \mathrm{or} \;\; e_k^c > E_c \\
      - \big[\frac{e_k(t)}{l_k(t)/l^{\mathrm{max}}_{k}}\big] & \text{otherwise} 
    \end{cases}
    \label{eq:reward}
  \end{equation}
where, in the first condition, we omit the dependence on $t$ of $\mathcal{M}_c$, $l_k$, $e_k^c$, and $E_c$ for simplicity. 

The reward is thus computed as the negative value of the ratio of the energy consumed $e_k(t)$ to the normalized latency $l_k(t) / l^{\mathrm{max}}_k$.  A 
penalty (i.e., a large constant number) is instead applied when any of the specified constraints are violated.

In MAPPO, the global state observation during the centralized training is formed by concatenating the local states of all agents. However, in LAPPO, we adopt an approach aiming at limiting the input size and improves performance \cite{NEURIPS2022}. We thus aggregate the total amount of $f_k$ requested by all cars connected to the same MEC node 
and include this global value as the extra information to the cars' local state observation to form a global state observation. This prevents increasing the input dimension unnecessarily with the number of cars since it introduces only a single input to the state, irrespective of the total number of cars.

\subsection{Mobility-Aware LAPPO (MALAPPO)}
In MALAPPO, we use the results of our empirical study in Sec.  \ref{sec:empriical_mob} to account for car mobility. Therefore, we adjust the state space of LAPPO by (i) adding a new parameter $mob_c$ presenting the daily mobility behavior of the car, and (ii) revising $\mathrm{map}_m$ to a continuous value between 0 and 1, thus indicating a \textit{connection preference} to \ac{MEC} node $m$. Incorporating car mobility (i.e., $mob_c$) into the state space is essential for better selecting the \ac{MEC} node for each car. For example, if the car is in the \textit{high mobile} cluster and the task latency requirement can still be met, it is preferable to select a higher tier \ac{MEC} node, which has a larger coverage and ensures the \ac{MEC} availability as the car moves. To compute the preference $\mathrm{map}_m$, we adapt the concept of \ac{MEC} node preference from \cite{Zhang2024}, which uses a weighted sum of the distance between the user and \ac{MEC} node, the available computation capacity at the node, and the communication latency. Instead of distance, we use the car's connection time to the \ac{MEC} node over the last 24 hours ($ct_m(t)$). As reported in Sec. \ref{sec:empriical_mob}, cars follow rather predictable mobility patterns and frequently connect to the same BSs and, thus, MEC nodes. By considering their connection time history, we can thus select a suitable node and avoid too frequent changes of \ac{MEC} nodes.
Thus, the preference map computes as $\mathrm{map}_m(t) = w_1 \cdot ct_m(t) + w_2 \cdot cpu_m(t) + w_3 \cdot [1/ l^{m, \mathrm{comm}}_{k}(t)]$.

In terms of complexity, we observe that, for both LAPPO and MALAPPO, it primarily involves processing the global state during the training phase, which is invariant to the number of agents (cars) since we aggregate the total request from all cars. The training complexity is thus a function of the number of training steps and the neural network architecture (e.g., number of layers). Decentralized execution, on the other hand, is computationally inexpensive as it primarily involves forward passes. We also observe that data privacy and security are not taken into account in this study, but our proposed model can be extended to include related mechanisms \cite{Li2022Privacy, Yu2023Privacy}.

\section{Performance Evaluation}
\label{sec:eval_main}
\subsection{Simulated Static Scenario}\label{sec:eval_static}

In this section, we first evaluate LAPPO in a simulated scenario with stationary users. We choose to use a simulation-based setup in order to cover a wider set of topologies (i.e. randomly generated) and highlight the energy efficiency that can be achieved by LAPPO compared to a set of baseline methods under various conditions. In the next section, we extend our analysis by testing MALAPPO in real-world scenarios.

We consider 10 \acp{BS}, and 4 \ac{MEC} nodes with randomly assigned computational capacity $\mathrm{CPU}_m \in [40, 60]$ GHz and maximum transmission rate $R_m \in \{10, 25, 100\}$ Gb/s \cite{shew2018transport}. We also assume cars with initial computational capacity $cpu_c \in [1, 2]$ GHz and initial battery energy level $E_c \in [50, 150]$ mJ. Cars are randomly placed in a 40$\times$40 km$^2$ area and, in each time slot, generate tasks with $data_k \in [200, 1000]$ Kbytes and $cl_k = 25$ cycles/bit. We assume latency thresholds for vehicular tasks $l^{\mathrm{max}}_k \in \{10, 30, 50, 100\}$ ms \cite{5gpp2015}. Moreover, $p^{\mathrm{tx}}_c = 0.5$ W and $p^{\mathrm{tx}}_b = 1$ W \cite{Sun2023, Ferdouse2019}. Unless specified otherwise, we consider $\alpha = 1$, i.e., car and MEC node energy consumption are given equal importance.

In LAPPO, each agent (car) is composed of two units, i.e., the actor and the critic, which are implemented as neural networks with the PyTorch framework. We consider architectures with two hidden layers of 128 neurons, Rectifier Linear Unit (ReLU) as the activation function, and  Adam Optimizer with a learning rate of $10^{-4}$. We also adopt common practices in implementing PPO, including Generalized Advantage Estimation and value-clipping of 0.2 \cite{Sun2023}. 

We compare LAPPO against four baselines:

\textbf{Random:} Tasks are randomly offloaded to \ac{MEC} nodes. 

\textbf{Local:} Tasks are computed locally by the cars. 

\textbf{Closest:} Tasks are offloaded to the closest \ac{MEC} nodes (in terms of minimum communication latency).

\textbf{HDMAPPO \cite{Sun2023}:} A MAPPO algorithm that aims at optimizing local energy consumption, latency, and drop rate for a 
task computation strategy. 
We modified HDMAPPO to optimize the total (car and MEC) energy consumption. 

\begin{figure}[t!]
\includegraphics[width=\linewidth]{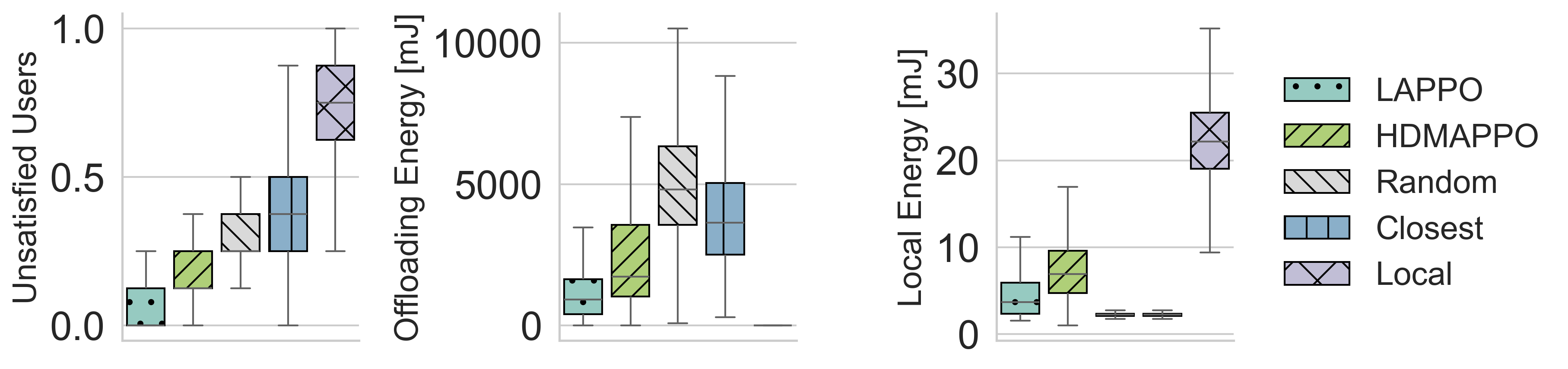}
    \caption{Ratio of unsatisfied users (left), offloading energy (center) and local energy (right) for LAPPO and baseline solutions in the static scenario.}
    \label{fig:time_performance}
    \vspace{-7mm}
\end{figure}

For simplicity, we first evaluate the performance with 8 cars, and later also analyze the impact of increasing the number of cars on the performance. Fig. \ref{fig:time_performance} shows the boxplot for three metrics, in which each point indicates the metric value at a time instant. Fig. \ref{fig:time_performance} (left) shows that LAPPO achieves the lowest rate of unsatisfied users (i.e., cars for which the task latency budget is exceeded) with a median of 0, while HDMAPPO has a median of 0.1. 'Local' performs the worst, with about 75\% of tasks exceeding their budget. 'Random' and 'Closest' show medians of 0.25 and 0.375, respectively. In Fig. \ref{fig:time_performance}, we also analyze total offloading energy $e_k^{\mathrm{off}}$ (center) and total local energy $e_k^{c}$ (right) across cars. LAPPO obtains the lowest offloading energy among all methods, except for 'Local', which does not offload any tasks and results in zero offloading energy. LAPPO reduces the offloading energy by 47\%, 75\%, and 81\% compared to HDAMPPO, 'Closest', and 'Random', respectively. In terms of local energy, 'Local' consumes the most energy as expected, while 'Random' and 'Closest' obtain the lowest consumption since they only rely on MEC offloading. LAPPO and HDMAPPO strike a balance between local and \ac{MEC} offloading, with LAPPO reducing the local energy consumption by 47\% compared to HDMAPPO.

\begin{figure}[t!]
\subfloat{
\includegraphics[width=0.5\linewidth]{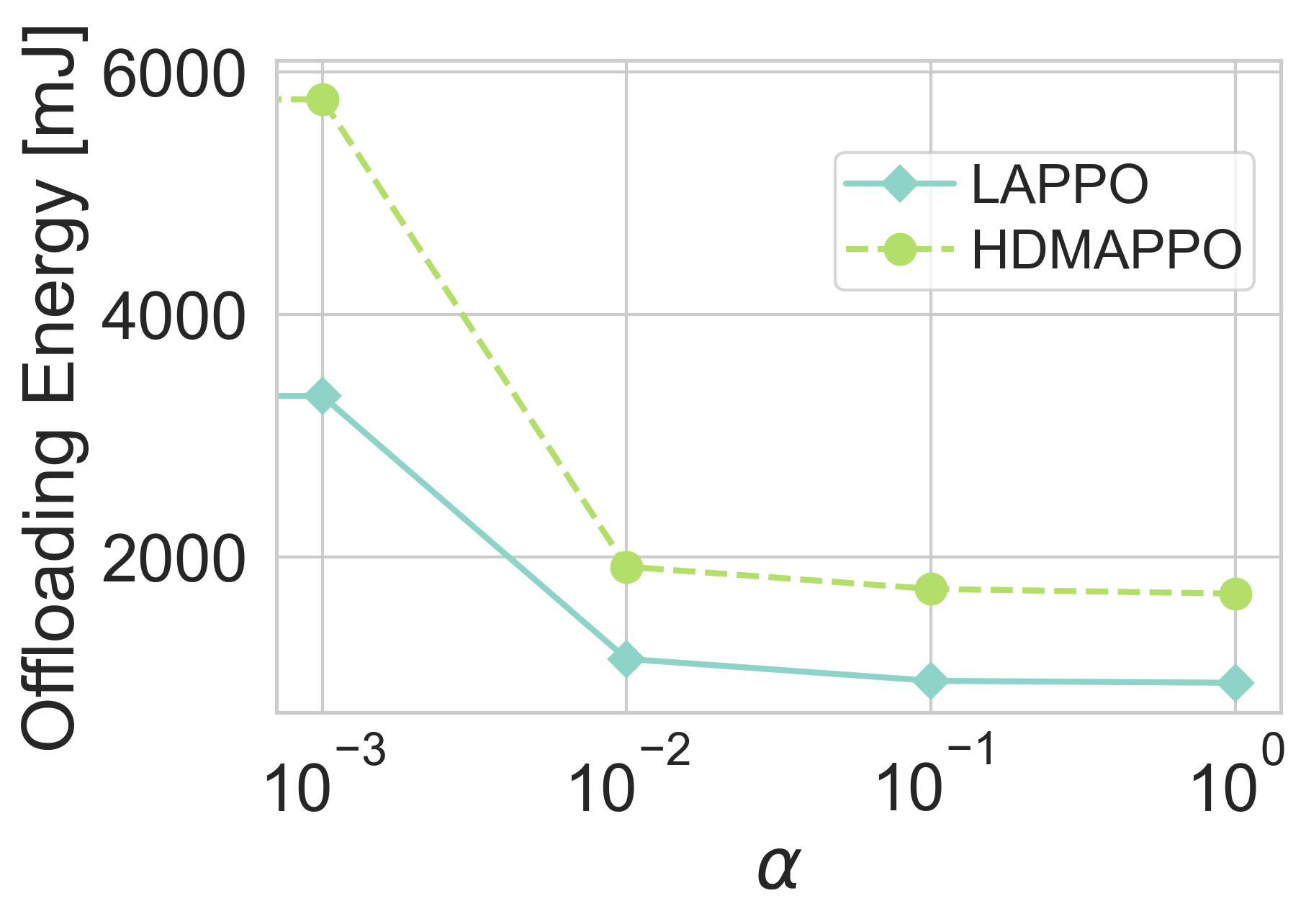}
    \label{fig:alpha_off_energy}
}
 \subfloat{
\includegraphics[width=0.46\linewidth]{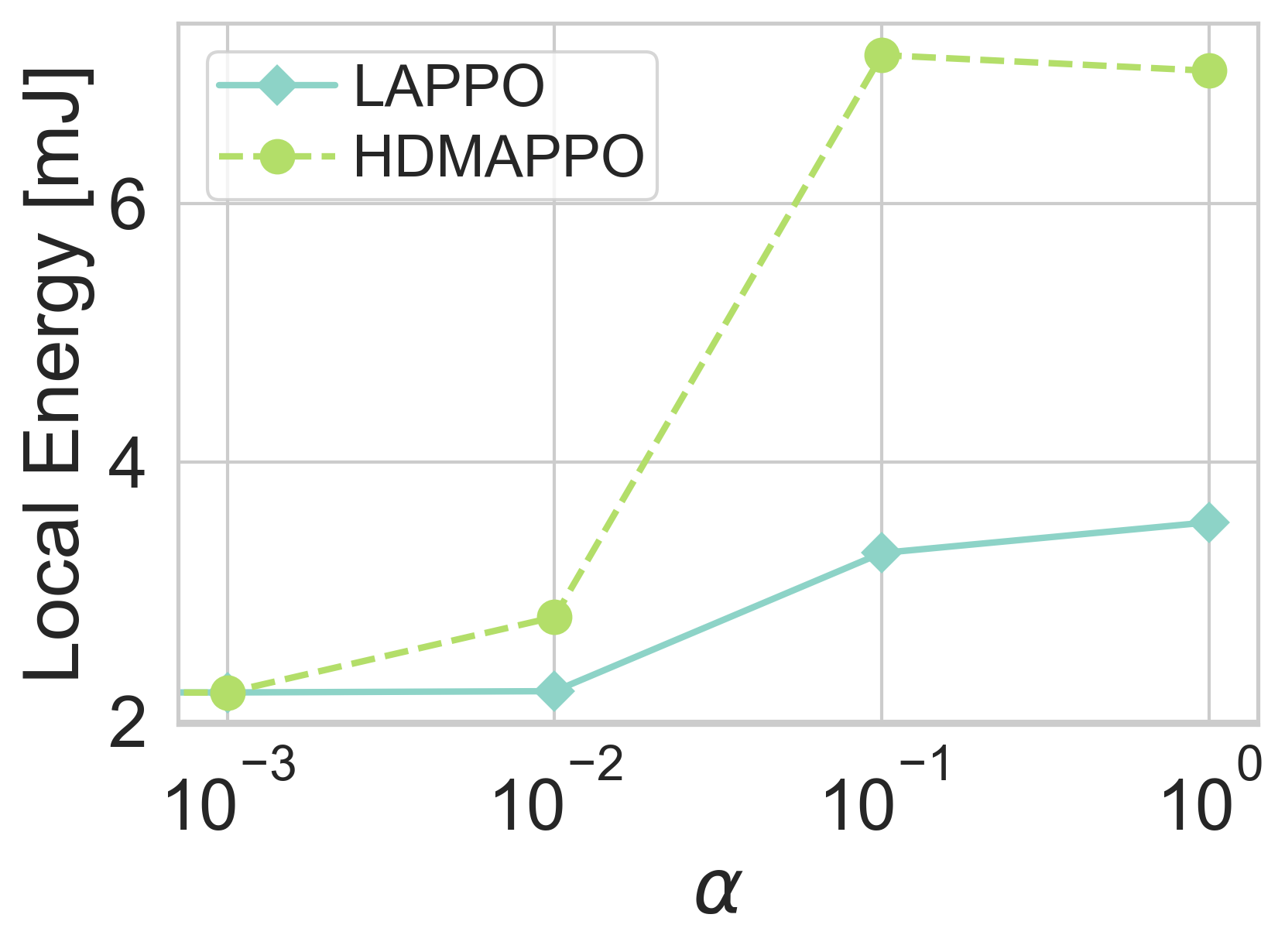}
    \label{fig:alpha_loc_energy}
}
\caption{Offloading (left) and local (right) energy consumption as a function of $\alpha$ in the static scenario.}
\label{fig:alpha_energy}
\end{figure}

\begin{figure}[t!]
\centering
\includegraphics[width=0.5\columnwidth]{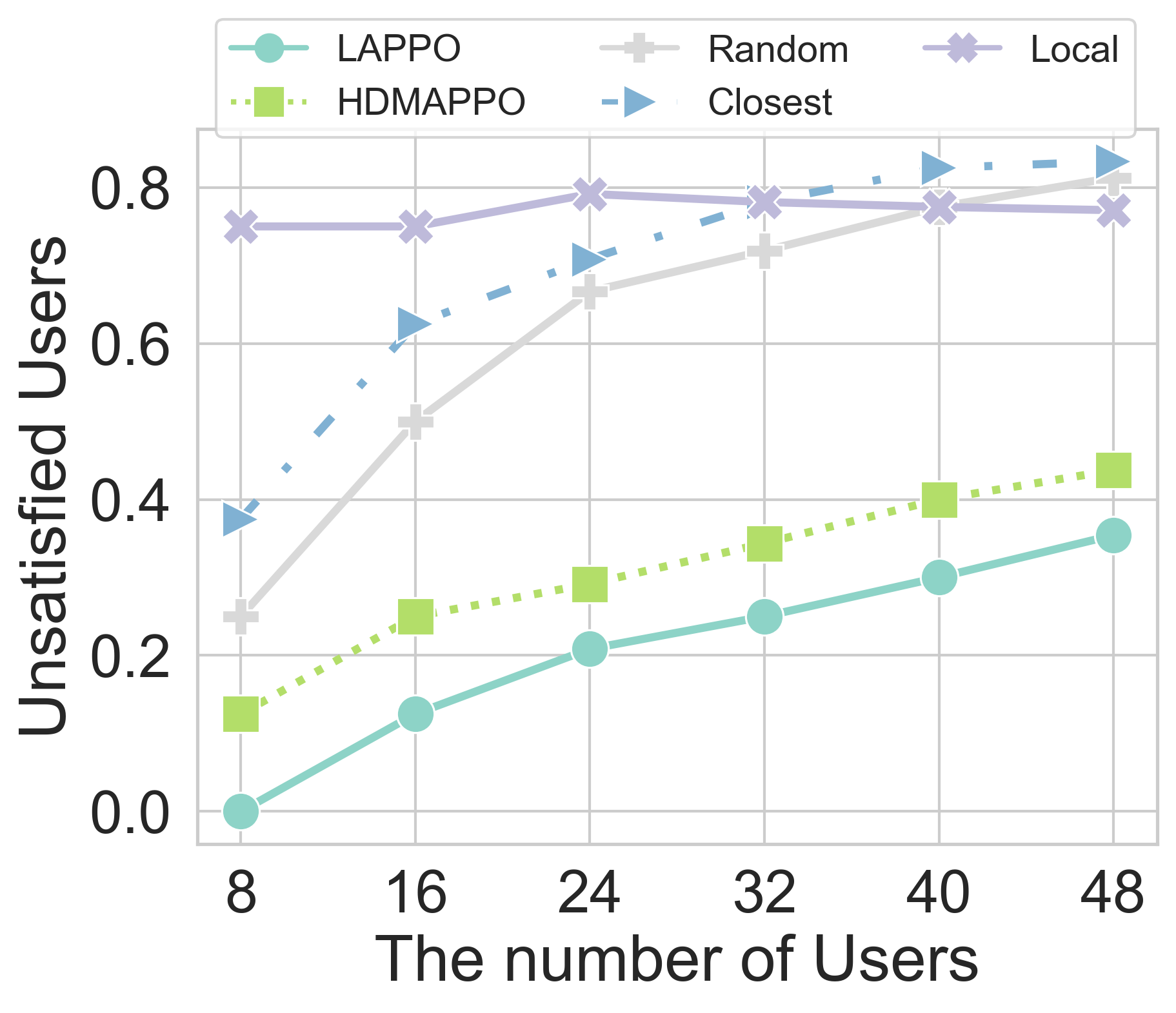}
\caption{Ratio of unsatisfied users as a function of the number of users in the static scenario.}
\vspace{-7mm}
\label{fig:droprate_numCars}
\end{figure}


Fig. \ref{fig:alpha_energy} shows the effect of changing $\alpha$ in \eqref{eq:e_k} on offloading (left) vs. local (right) energy consumption for LAPPO and HDMAPPO. 
By taking the median over 1000 iterations, we observe that, as $\alpha$ increases from 0 to 1, the offloading energy reduces by 76\% while local energy rises by 57\% for LAPPO. This confirms that an exclusive focus on minimizing the local energy (i.e., $\alpha \simeq 0$) results in high offloading energy consumption. Additionally, LAPPO outperforms HDMAPPO with a 29\% ($\alpha = 0$) to 43\% ($\alpha = 1$) lower offloading energy.

In Fig. \ref{fig:droprate_numCars}, we study how the performance of LAPPO and the baseline algorithms change as we increase the number of cars while MEC resources remain limited. As expected, the rate of unsatisfied users increase for all methods with growing number of cars, except for 'Local'. This is due to the higher competition for MEC resources, given that the MEC computational resources remain unchanged. 'Local' is not affected since cars use only their own local resources. LAPPO and HDMAPPO outperform 'Random' and 'Closest', with LAPPO reducing the unsatisfied users by 20\% compared to HDMAPPO when the number of cars is 48, showcasing higher robustness in crowded scenarios.

As a final analysis for the static scenario, we use the Gurobi solver \cite{Gurobi} to compute the optimal solution for the optimization problem in Sec. \ref{sec:opt_model}. The task computation problem is a typical generalized assignment problem (GAP), thus the computational requirements for solving it rise rapidly, even with a slight increase in the number of users \cite{ROMEIJN2000}. Therefore, for this analysis, we focus on small scenarios (8 and 16 cars) as these still provide the ground for representative results. 
The optimal solution assumes a single centralized node with complete information about all cars to allocate their tasks, which is rather unpractical in real-world scenarios. We compare LAPPO with this optimal 
solution for scenarios with 8 and 16 cars. For 8 cars, the median total energy of LAPPO is 36\% higher. For 16 cars, the median total energy of LAPPO is 1\% lower, but LAPPO results in 10\% of unsatisfied users (median over 10000 iterations), while the optimal solution keeps the rate of unsatisfied users equal to 0\%. Therefore, despite LAPPO being a distributed solution with partial information observability, it demonstrates performance comparable to the optimal solution for the scenarios targeted in this paper.

\subsection{Real-world Mobile Scenario}\label{sec:eval_mob}

In this section, we evaluate MALAPPO under real-world settings, i.e., the network topology and car connection/mobility information in various scenarios observed for the \ac{MNO} under study.  
Following Fig. \ref{fig:problem_statement}, we assume a 3-tier \ac{MEC} deployment consisting of 26 Tier-1, 1 Tier-2, and 1 Tier-3 \ac{MEC} nodes, for a total covered area of about 157$\times$192 km$^2$. Based on Intel guidelines for edge applications \cite{intel_iot_edge}, we set the computational capacity of \ac{MEC} nodes in Tier-1 as 78 GHz, in Tier-2 as 131 GHz, in Tier-3 as 187 GHz. We also consider $w_1 = 0.4, w_2 = 0.3, w_3 = 0.3$ to compute $map_m$ in MALAPPO.
Other parameters align with those in the Section \ref{sec:eval_static}. We assume that cars obtain information on the available \ac{MEC} nodes from their connected \ac{BS} at each time. Each BS has the updated states of their accessible \ac{MEC} nodes, limited by the underlying network topology.  When a car moves, connects to a new \ac{BS}, and finds its previously selected \ac{MEC} node unavailable (out of coverage), it selects a new \ac{MEC} node or starts a local computation. If the task was not completed in the previous time slot, the car restarts the task computation from scratch, because tasks are atomic and cannot be split.

We compare MALAPPO against LAPPO and two state-of-the-art mobility-aware algorithms: 

\textbf{MWBS \cite{Zhang2024}:} This algorithm calculates the preference of a car towards a \ac{MEC} node based on a weighted sum of three factors: distance and communication latency between car and node, and available computational resources at the node. Unlike the original study, which classifies cars into high- and low- mobile groups based on speed and assigns each group a preference model, we leverage clustering results from Sec. \ref{sec:empriical_mob} to categorize cars based on their mobility patterns into high- and low-mobile groups.
We then use this preference as $\mathrm{map}_m$ in the state space of LAPPO.

\textbf{Next Location (NextLoc):} Some works (e.g., \cite{Cao2024, Li2019, Wei2020}) have attempted to predict user mobility using models or data-driven approaches to make task computation decisions. Rather than predicting the next location, we consider an ideal method where we know and use the actual next location.
We then compute $\mathrm{map}_m$ in the state space of LAPPO based on the intersection of \ac{MEC} nodes associated with both the current and the next \ac{BS} the car is connected. 

\begin{figure}[t!]
\subfloat{
\includegraphics[width=0.48\linewidth]{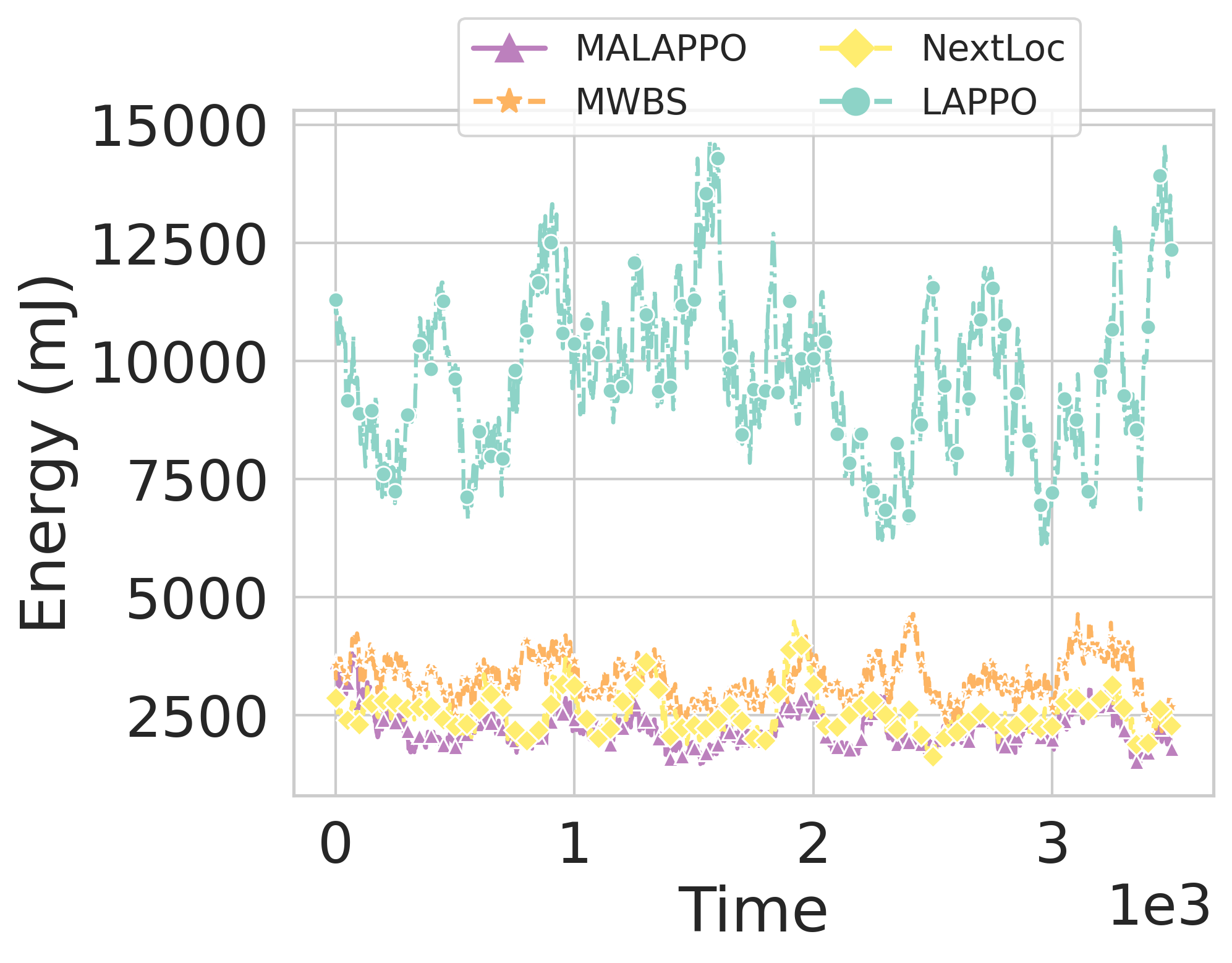}
    \label{fig:mobile_scenario_energy}
}
 \subfloat{
\includegraphics[width=0.46\linewidth]{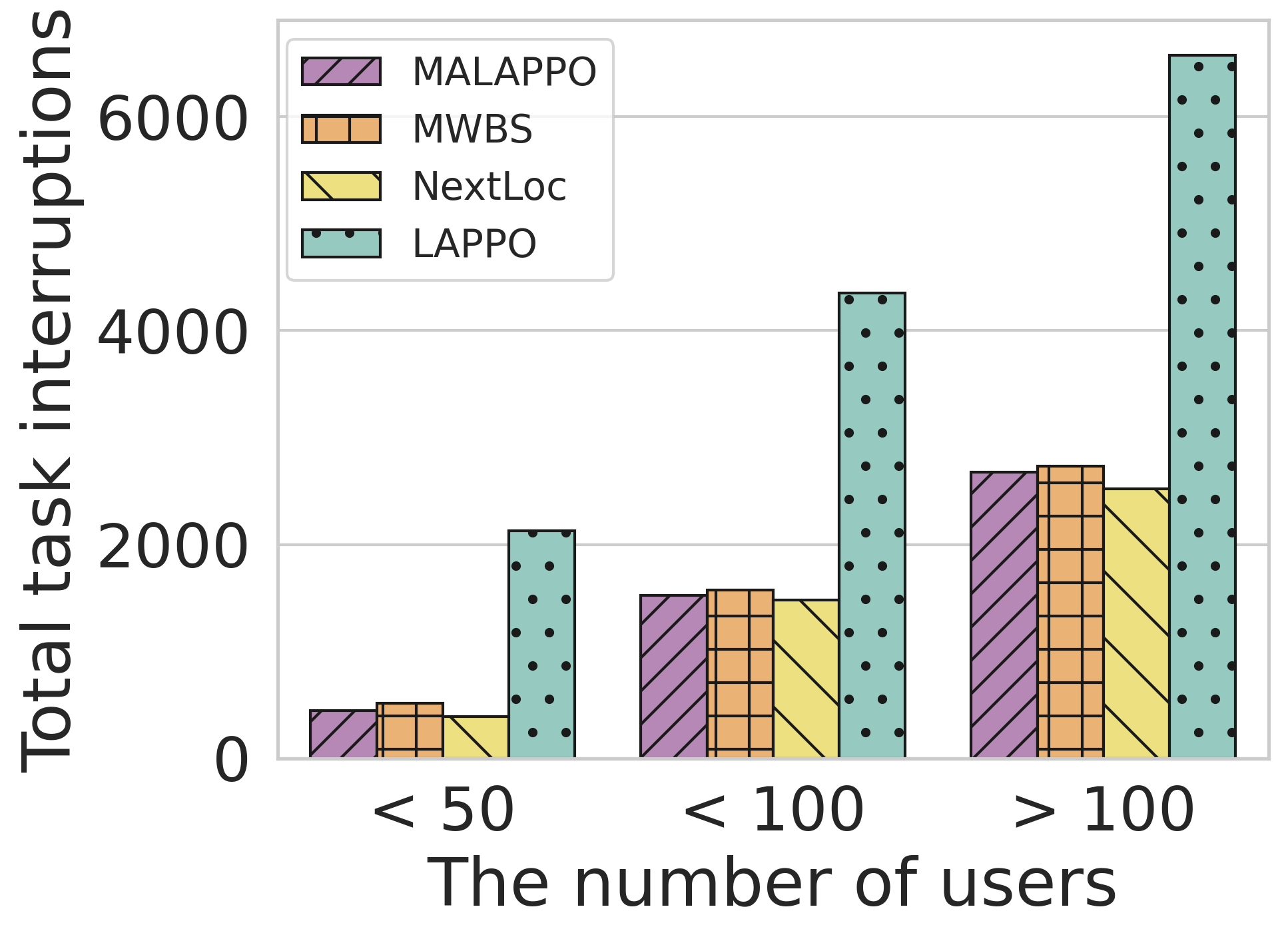}
    \label{fig:mobile_scenario_interruptions}
}
\caption{Total energy consumption over time (left) and number of task interruptions as a function of the number of users (right) for MALAPPO and baseline solutions in the mobile scenario.}
\vspace{-6mm}
\label{fig:mobile_scenario}
\end{figure}

Fig. \ref{fig:mobile_scenario} (left) shows the total energy consumption of the various methods over time (after training). To enhance readability, each point in the figure represents the moving average over 100 time slots. The results show that LAPPO consumes more energy than the other mobility-aware methods, highlighting the inefficiency of non-mobility-aware approaches. This is due to the frequent reallocation and restarting of tasks in a new \ac{MEC} node. Conversely, MALAPPO outperforms all the baselines, reducing the total energy consumption by 78\%, 27\%, and 14\% compared to LAPPO, NextLoc, and MWBS, respectively. 
Fig. \ref{fig:mobile_scenario} (right) shows the total number of task interruptions due to the car mobility. While LAPPO experiences a high rate of task interruptions, all mobility-aware methods decrease task interruptions. MALAPPO and MWBS perform similarly to the NextLoc, which benefits from knowing the next car location. However, accurately predicting a car's location requires extensive data, potentially including sensitive information like historical Global Positioning System (GPS) coordinates. In scenarios with more than 100 concurrent cars, MALAPPO shows $\sim$6\% higher interruptions compared to NextLoc.

\section{Conclusion}\label{sec:conclusion}

In this paper, we propose an energy-efficient task computation strategy for V2X services. To do so, we first carry out a car mobility analysis from a real-world dataset. Our analysis shows that incorporating both stationary and mobility-aware solutions is crucial for optimizing task computation, aiming at ensuring user satisfaction and system energy efficiency under various mobility scenarios. Hence, we design a solution to adapt to both static and mobile scenarios, employing a decentralized PPO-based algorithm. 
Our solution shows significant energy savings of 47\% in static scenarios and reduces user dissatisfaction by 20\% in these settings. Additionally, our mobility-aware solution can decrease energy consumption by 14\% in the mobile scenarios compared to state-of-the-art schemes. 
In future work, we plan to enhance our mobility model by analyzing mobility features over varying time scales to identify the best temporal granularity for our model to remain efficient across various scenarios. Furthermore, given the critical role of privacy and security in vehicular networks, we intend to investigate how they can be integrated into our solution to ensure secure task computation. 
\section*{Acknowledgements}

 This work by P.P. and O.A is partially supported by the EU HE programme (agreement No.
101096452, IMAGINE-B5G). The work by G.C. is supported by the Knowledge Foundation of Sweden and by the EU HE programme (agreement No. 101139172, 6G-PATH). The work by J.I. is partially supported by UNICO I+D Cloud program with the CLOUDLESS project and the EDGELESS project, funded by the EU HE programme (agreement No. 101092950).

\bibliographystyle{IEEEtran}
\bibliography{ref}

\end{document}